\documentstyle[12pt]{article}
\begin{document}
\newcommand{\ba}{\begin{array}}
\newcommand{\ea}{\end{array}}
\newcommand{\eee}{\mbox{e}}
\newcommand{\ovl}{\overline}
\newcommand{\pal}{\partial}
\newcommand{\triex}{\mbox{\hspace{3ex}}}
\newcommand{\edoc}{\end{document}}
\newcommand{\Pfaff}{\mbox{Pfaff}}
\newcommand{\sumsigma}{\ba[t]{c} \mbox{$\Sigma$} \vspace{-1ex} \cr
  \mbox{$\scriptstyle{\{\,N\,\}}$} \ea }
\newcommand{\spsigma}{\ba[t]{c} \mbox{Sp} \vspace{-1ex} \cr
  \mbox{$\scriptstyle{(\sigma)}$} \ea }
\newcommand{\spsigmamn}{\ba[t]{c} \mbox{Sp} \vspace{-1ex} \cr
  \mbox{$\scriptstyle{(\sigma_{mn})}$} \ea }
\newcommand{\spura}{\ba[t]{c} \mbox{Sp} \vspace{-1ex} \cr
  \mbox{$\scriptstyle{(a)}$} \ea }
\newcommand{\spurasigma}{\ba[t]{c} \mbox{Sp} \vspace{-1ex} \cr
  \mbox{$\scriptstyle{(\sigma\,|\,a)}$} \ea }

\thispagestyle{empty}

\begin{center}
{\bf ANTICOMMUTING INTEGRALS AND FERMIONIC FIELD THEORIES \\
FOR TWO-DIMENSIONAL ISING MODELS} $\dagger$  \vspace*{3ex} \\
                   V.N. Plechko $\ddagger$
\vspace*{2ex} \\
{\em Bogoliubov Theoretical Laboratory, Joint Institute for
Nuclear Research, \\ JINR-Dubna, 141980 Dubna, Moscow Region,
Russian Federation}
\vspace*{5ex} \\

     {\bf Abstract} \end{center}

\noindent  We review the applications of the integral over anticommuting
Grassmann variables (nonquantum fermionic fields) to the analytic solutions
and the field-theoretical formulations for the 2D Ising models. The 2D
Ising model partition function $Q$ is presentable as the fermionic Gaussian
integral. The use of the spin-polynomial interpretation of the 2D Ising
problem is stressed, in particular. Starting with the spin-polynomial
interpretation of the local Boltzmann weights, the Gaussian integral for
$Q$ appears in the universal form for a variety of lattices, including the
standard rectangular, triangular, and hexagonal lattices, and with the
minimal number of fermionic variables (two per site). The analytic
solutions for the correspondent 2D Ising models then follow by passing to
the momentum space on a lattice. The symmetries and the question on the
location of critical point have an interesting interpretation within this
spin-polynomial formulation of the problem. From the exact lattice theory
we then pass to the continuum-limit field-theoretical interpretation of the
2D Ising models. The continuum theory captures all relevant features of the
original models near $T_c$. The continuum limit corresponds to the
low-momentum sector of the exact theory responsible for the critical-point
singularities and the large-distance behaviour of correlations. The
resulting field theory is the massive two-component Majorana theory, with
mass vanishing at $T_c$. By doubling of fermions in the Majorana
representation, we obtain as well the 2D Dirac field theory of charged
fermions for 2D Ising models. The differences between particular 2D Ising
lattices are merely adsorbed, in the field-theoretical formulation, in the
definition of the effective mass. \\

\noindent $\dagger$\
Proceedings of the VII International Conference on Symmetry Methods in
Physics, Dubna, Russia, ICSMP-95, July 10--16, 1995.  Ed. by A.N. Sissakian
and G.S. Pogosyan (Publ. JINR E2-96-224, Dubna, 1996)  Vol.~2,
p.~443--450.\\

\noindent $\ddagger$\ E-mails:
plechko@thsun1.jinr.dubna.su; \
plechko@theor.jinrc.dubna.su \\*[3ex]

\noindent{\large  hep-th/9607053  }  \\*[-1ex]


\newpage\setcounter{page}{1}

       \begin{center}
    {\bf ANTICOMMUTING INTEGRALS AND FERMIONIC FIELD THEORIES \\
         FOR TWO-DIMENSIONAL ISING MODELS }  \vspace*{2ex} \\
                      V.N. Plechko
       \vspace*{1ex} \\
{\em Bogoliubov Theoretical Laboratory, Joint Institute for
Nuclear Research, \\ JINR-Dubna, 141980 Dubna, Moscow Region,
Russian Federation \\ e-mail: plechko@theor.jinrc.dubna.su}
        \end{center}

     \begin{abstract}
\noindent  We review some aspects of the anticommuting integrals as applied
to the analytic solutions and the field-theoretical formulations for the 2D
Ising models. We stress, in particular, the use of the spin-polynomial
interpretation of the 2D Ising problem.
     \end{abstract}

{\em 1. Introduction.}\ A remarkable feature of the two-dimensional
Ising model, first established within the transfer-matrix and combinatorial
considerations, is that it can be reformulated as a free fermion field
theory. It was recognized later on that the natural tool to analyze the
$2D$ Ising models is the notion of the integral over Grassmann variables
(nonquantum fermionic fields) introduced by Berezin [1-8]. A simple
interpretation of the $2D$ Ising model based on the integration over
anticommuting Grassmann variables and the mirror-ordered factorization
principle for the density matrix was introduced in [7,8]. The partition
function is expressible here as a fermionic Gaussian integral even for the
most general inhomogeneous distribution of the coupling parameters [7].
This also gives a few line derivation of the Onsager result [7]. The
Gaussian fermionic representations has been recently constructed as well
for the inhomogeneous dimer problems [9]. The method [7,8] do not involve
traditional transfer-matrix or combinatorial considerations.
Schematically, we have:
$$
\ba{llr}
Q\,=\,\spsigma\,Q\,(\sigma)\to\spurasigma\,Q\,(\sigma\,|\,a)\to
\spura\,Q\,(a)\,=\,Q\,.
\ea\eqno(1)
$$
We start here with the original spin partition function $Q$, and then
we introduce, in a special way, a set of new anticommuting Grassmann
variables $(a)$, thus passing to a mixed $(\sigma\,|\,a)$ representation.
Eliminating spin variables in this mixed representation, we obtain purely
fermionic expression (fermionic Gaussian integral) for the same partition
function $Q$. Grassmann variables are introduced, at the first stages, by
factorization of the local weights. The factorization ideas resemble, in a
sense, the idea of insertion of Dirac's unity, $\Sigma\,|a\left > \right <
a|=1$, in transformations in quantum mechanics [10]. An important
ingredient of the method is also the mirror-ordering procedure for the
arising noncommuting Grassmann factors [7,8]. The original variant of the
method implies two Grassmann variables {\em per bond} \ in the resulting
fermionic integral, since we start here with the factorization of the bond
Boltzmann weights [7]. An important modification was introduced in [8],
where we start with the factorization of the {\em cell} \ weights presented
by three-spin polynomials. This enables us to obtain the fermionic
representation for $Q$ with only two fermionic variables {\em per site} \
which provides essential simplifications in the analysis [8]. The
symmetries and the criticality conditions have interesting spin-polynomial
interpretation [8]. In this report we comment on the analytic results
obtained so far within this spin-polynomial fermionic approach and derive
the effective field-theoretical (continuum-limit) formulations for the 2D
Ising models valid near $T_c$. The resulting field theory is the massive
free-fermion Majorana theory. The form of the action is universal,
particular Ising lattices, like rectangular, triangular and hexagonal
lattices, are specified merely by the definition of mass. By doubling of
fermions in the Majorana representation, we obtain as well the 2D Dirac
field interpretation for 2D Ising models near $T_c$.

{\em 2. Free-fermion representation and analytic results (lattice case).}

The spin-polynomial interpretation of the 2D Ising models can be
illustrated by an example of the triangular lattice [8]. Let us consider
the triangular Ising lattice as a rectangular one with the additional
diagonal interaction introduced in each lattice cell. Let $mn$ be the
lattice sites on the corresponding rectangular net, $m,n=1,2,3,..., L$,
where $L$ is lattice length, $N=L^2$ is the number of lattice sites. At
final stages we assume $L^2\to \infty$. The hamiltonian of the triangular
2D Ising model is [8]:
$$
\ba{llr}
-\beta\,H\,(\sigma)=\sum\limits_{mn}^{}\;[\;b_1\sigma_{mn}
\sigma_{m+1n}+ b_2\sigma_{m+1n}\sigma_{m+1n+1}+ b_3\sigma_{mn}
\sigma_{m+1n+1}\,]\,, \;\;\; \beta=1/kT\,,
\cr\ea\eqno(2)
$$
where $\sigma_{mn}=\pm1$ are the Ising spins, $b_{\alpha} = J_{\alpha}/
kT$ are dimensionless coupling constants, $J_{\alpha}$ are the exchange
energies, $kT$ is temperature in energy units. In what follows we assume
the purely ferromagnetic case, $b_{\alpha}>0$. Respectively, the
partition function and the free energy per site are given as follows:
$$
\ba{llr}
Z=\sum_{(\sigma)}^{}\,\eee^{\,-\beta H(\sigma)}\,,\triex
-\beta\,f_{\,Z}=\, \lim\limits_{N\to\infty}^{}\frac{1}{N}\ln\,Z\,,
\ea\eqno(3)
$$
where the sum is taken over $2^N$ spin configurations provided by
$\sigma_{mn}=\pm1$ at each site. Noting the identity for the typical
bond weight $\eee^{\,b\,\sigma\sigma'}=\cosh\,b\,+\,\sinh\,b\cdot\sigma
\sigma'$, which readily follows from $(\sigma\sigma')^{2}=+1$, we obtain:
$$
\ba{llr}
Z=\,(2\,\cosh\,b_1\,\cosh\,b_2\,\cosh\,b_3)^{\,N}\,Q\,, \triex
N=L^{2}\to\infty\,,
\ea\eqno(4)
$$
where $Q$ is the reduced partition function [8]:
$$
\ba{llr}
Q = \spsigma\{\,\prod\limits_{mn}\,(\,1+t_1\sigma_{mn}\sigma_{m+1n})\,
(1+t_2\sigma_{m+1n}\sigma_{m+1n+1})\,(1+t_3\sigma_{mn}
\sigma_{m+1n+1})\,\}\,,
\ea\eqno(5)
$$
where $t_{\alpha}= \tanh\, b_{\alpha}$, and $Sp_{(\sigma)}=
2^{\,-N}\Sigma_{ (\sigma)}$ is the normalized spin averaging.

The spin-polynomial interpretation arises if we multiply the three bond
weights forming a triangular cell in $Q$. Noting that $\sigma_1 \sigma_2
\cdot \sigma_2\sigma_3 = \sigma_1 \sigma_3$, etc, we obtain [8]:
$$
\ba{llr}
Q=\spsigma \prod\limits_{mn}^{}\,(\alpha_0 + \alpha_1\,
\sigma_{mn}\sigma_{m+1n}+ \alpha_2\,\sigma_{m+1n}\sigma_{m+1n+1}+
\alpha_3\,\sigma_{mn}\sigma_{m+1\,n+1}\,)\,,
\cr\ea\eqno(6)
$$
where for the triangular lattice we have $\alpha_0=1+t_1t_2t_3\,,\,
\alpha_1= t_1+ t_2t_3\,,\, \alpha_2=t_2+ t_1t_3\,,\, \alpha_3=t_3+
t_1t_2\,$, the case of the rectangular lattice follows with $t_3=0$.  As is
discussed in [8] the hexagonal and some other lattices are covered by (6)
as well, with the correspondent specification of the $\alpha$ parameters.
In what follows we consider the three-spin polynomial partition function
$Q$, assuming that $\alpha_0, \alpha_1, \alpha_2,\alpha_3\,$ are some
numerical parameters.

The partition function (6) can be transformed into the fermionic
Gaussian integral as follows [8]:
$$
\ba{llr}
Q\, =
\int \prod\limits_{mn}^{}dc_{mn}^{*}dc_{mn}^{}\exp\,\{
\sum\limits_{mn}^{}\,[\,(\alpha_{0}\,c_{mn}c_{mn}^{*} -
\alpha_1\,c_{mn}c_{m-1n}^{*} - \alpha_2\,c_{mn}c_{mn-1}^{*}-\cr
-\,\alpha_{3}\,c_{mn}c_{m-1n-1}^{*}) -\alpha_{1}\,c_{mn}c_{m-1n}-\,
\alpha_{2}\,c_{mn}^{*}c_{mn-1}^{*}\,]\,\}\;,
\cr\ea\eqno(7)
$$
where $c_{mn}^{}, c_{mn}^{*}$ is a set of purely anticommuting Grassmann
variables. The free-fermion representation for $Q$ given in (7) is
exact and completely equivalent to the original spin representation (6) up
to the boundary effects negligible in the limit of infinite lattice
$L^2\to\infty$. [We do not comment here neither on the properties of the
Grassmann variables, nor on the method of the derivation of this
representation. The derivation is  simple and takes one page of
transformations, neglecting preliminary preparations, the details are to be
seen in [8]. Also see the comments on the fermionic Gaussian integrals in
[9] ].

The explicit evaluation of the integral (7) can be performed by passing to
the momentum space. For the squared partition function we find [8]:
$$
\ba{llr}
Q^{\,2} = \prod\limits_{p=0}^{L-1}\prod\limits_{q=0}^{L-1}\,[\,
(\alpha_{0}^{2}+\alpha_{1}^{2}+\alpha_{2}^{2}+\alpha_{3}^{2}) -
2\,(\alpha_{0}^{}\alpha_{1}^{}-\alpha_{2}^{}\alpha_{3}^{})\,\cos
\frac{2\pi p}{L}\, - \cr -\,
2\,(\alpha_{0}^{}\alpha_{2}^{}-\alpha_{1}^{}\alpha_{3}^{})\,\cos
\frac{2\pi q}{L} -
2\,(\alpha_{0}^{}\alpha_{3}^{}-\alpha_{1}^{}\alpha_{2}^{})\,\cos
\frac{2\pi (p+q)}{L}\,]\;.
\ea\eqno(8)
$$
Respectively, for the free energy we obtain:
$$
\ba{llr}
-\beta f_Q = \,\lim_{L\rightarrow\infty}^{}\,
(\frac{1}{L^2}\ln Q\,)\,
=\cr =  \frac{1}{2}\,\int\limits_{0}^{2\pi}\int\limits_{0}^{2\pi}
\frac{dp}{2\pi} \frac{dq}{2\pi}\,\ln\,[\,
(\alpha_{0}^{2}+\alpha_{1}^{2}+\alpha_{2}^{2}+\alpha_{3}^{2}) -
2\,(\alpha_{0}^{}\alpha_{1}^{}-\alpha_{2}^{}\alpha_{3}^{})\,\cos p\,-
\cr -\,
2\,(\alpha_{0}^{}\alpha_{2}^{}-\alpha_{1}^{}\alpha_{3}^{})\,\cos q\,-
2\,(\alpha_{0}^{}\alpha_{3}^{}-\alpha_{1}^{}\alpha_{2}^{})\,\cos (p+q)\,]\;.
\cr\ea\eqno(9)
$$
It is easy to check that the free-energy form (9) yields the known exact
expressions for the free energies of the rectangular, triangular, and
hexagonal Ising lattices, with the correspondent specifications of the
parameters $\alpha_{_j}$.

{\em 3. The symmetries and the critical point. }

The symmetries provided by the above exact solution and the question on the
location of the critical point have an interesting interpretation in the
spin-polynomial language [8]. Introducing the notation $(\sigma_{1},
\sigma_{2}, \sigma_{3}) _{mn} \to (\sigma_{mn}, \sigma_{m+1n},
\sigma_{m+1\, n+1} )$, the cell weight in the density matrix in (6) is
given as the three-spin polynomial:
$$
\ba{llr}
P_{\,123}\,(\sigma)= \alpha_0 +\alpha_1\,\sigma_1\sigma_2 +
\alpha_2\,\sigma_2\sigma_3 + \alpha_3\,\sigma_1\sigma_3\,.
\cr\ea\eqno(10)
$$
It is also useful to introduce the associated three-spin polynomial:
$$
\ba{llr}
F_{\,123}\,(\sigma)= \alpha_0 - \alpha_1\,\sigma_1\sigma_2 -
\alpha_2\,\sigma_2\sigma_3 - \alpha_3\,\sigma_1\sigma_3\,,
\cr\ea\eqno(11)
$$
and we note the following interesting identity [8]:
$$
\ba{llr}
(F_{\,123}(\sigma))^{\,2} = (\alpha_0 - \alpha_1\,\sigma_1\sigma_2 -
\alpha_2\,\sigma_2\sigma_3 - \alpha_3\,\sigma_1\sigma_3\,)^{\,2} = \cr
=\,(\alpha_{0}^{2}+ \alpha_{1}^{2}+\alpha_{2}^{2}+\alpha_{3}^{2})\,-\,
2\,(\alpha_0\alpha_1-\alpha_2\alpha_3)\,\sigma_1\sigma_2\,-\,\cr\,-\,
2\,(\alpha_0\alpha_2-\alpha_1\alpha_3)\,\sigma_2\sigma_3\,-\,
2\,(\alpha_{0}\alpha_{3}-\alpha_{1}\alpha_{2})\,\sigma_1\sigma_3\,.
\ea\eqno(12)
$$
It is seen that the combinations of the $\alpha$-parameters occurring in
$(F_{123})^2$ are exactly the same as in the momentum modes $Q^{\,2}(p\,|
\,q)$ in (9):
$$
\ba{llr} Q^{\,2}\,(p\,|\,q) =
(\alpha_{0}^{2}+\alpha_{1}^{2}+\alpha_{2}^{2}+\alpha_{3}^{2}) -
2\,(\alpha_{0}^{}\alpha_{1}^{}-\alpha_{2}^{}\alpha_{3}^{})\,\cos p\,-
\cr -\,
2\,(\alpha_{0}^{}\alpha_{2}^{}-\alpha_{1}^{}\alpha_{3}^{})\,\cos q\,-
2\,(\alpha_{0}^{}\alpha_{3}^{}-\alpha_{1}^{}\alpha_{2}^{})\,\cos\,(p+q)\,,
\ea\eqno(13)
$$
where $0\leq p,q \leq2\pi$, the limit $L^2\to\infty$ is already assumed.

It appears that the following combinations of the $\alpha$-parameters play
important role in discussing the symmetries and the critical point [8]:
$$
\ba{ll} \ba{llr}
\alpha_{0}^{*}=\frac{1}{2}\,(\alpha_0+\alpha_1+\alpha_2+\alpha_3)\,,\cr
\alpha_{1}^{*}=\frac{1}{2}\,(\alpha_0+\alpha_1-\alpha_2-\alpha_3)\,,\cr
\alpha_{2}^{*}=\frac{1}{2}\,(\alpha_0-\alpha_1+\alpha_2-\alpha_3)\,,\cr
\alpha_{3}^{*}=\frac{1}{2}\,(\alpha_0-\alpha_1-\alpha_2+\alpha_3)\,,
\ea & \;\;\; \ba{llr}
\bar{\alpha}_{0}^{}=\frac{1}{2}\,(\alpha_0-\alpha_1-\alpha_2-\alpha_3)\,,\cr
\bar{\alpha}_{1}^{}=\frac{1}{2}\,(\alpha_0-\alpha_1+\alpha_2+\alpha_3)\,,\cr
\bar{\alpha}_{2}^{}=\frac{1}{2}\,(\alpha_0+\alpha_1-\alpha_2+\alpha_3)\,,\cr
\bar{\alpha}_{3}^{}=\frac{1}{2}\,(\alpha_0+\alpha_1+\alpha_2-\alpha_3)\,.
\ea  \cr\ea\eqno(14)
$$
The parameters $\alpha^*$ and $\bar{\alpha}$ are in fact the eigenvalues
of the polynomials $\frac{1}{2}P_{123}$ and $\frac{1}{2} F_{123}$,
respectively. By the "eigenvalues" we mean the four numbers which takes the
polynomial as the spin variables run over their permissible values $\pm1$.
We note also the following identity:
$$
\ba{llr}
\bar{\alpha}_0\bar{\alpha}_1\bar{\alpha}_2\bar{\alpha}_3=
\alpha_{0}^{*}\alpha_{1}^{*}\alpha_{2}^{*}\alpha_{3}^{*}\,-\,
\alpha_0\alpha_1\alpha_2\alpha_3\,.
\cr\ea\eqno(15)
$$
There are some evident symmetries in the solution. For instance, the free
energy (9) is a symmetric function with respect to arbitrary permutations
of $\alpha_0, \alpha_1, \alpha_2,\alpha_3$ parameters. We can as well
change the signs of any two of them, with $-\beta f_Q$ unchanged. There is
also a less evident hidden symmetry in the solution, corresponding to the
Kramers-Wannier duality in the case of the standard rectangular lattice.
Namely, the partition function $Q\{\alpha_{}\}$ is invariant under the
transformation $\alpha_{j} \leftrightarrow \alpha_{j}^ {\,*}$. This
symmetry in fact holds already for the parameters of the separable
fermionic modes (13), and can be proved making use of (12), see [8].

In order to establish the possible critical points, we have to look for
zeroes of $Q^{\,2}\,(p\,|\,q)$ momentum modes. As it can be guessed
already from the analogy between (12) and (13), there are four exceptional
modes with $(p\,|\,q)=\, (0\,|\,0), (0\,|\,\pi), (\pi\,|\,0), (\pi\,|\,
\pi)$. For these modes we have, respectively, $Q^{\,2}\, (p\,|\,q)=
(2\bar{\alpha}_0)^2\,,  \;(2\bar{\alpha}_1)^2\,,\; (2\bar{\alpha}_2)^2\,,
\;(2\bar{\alpha}_3)^2\,$. We remember that the parameters $\alpha_{j}^{}$
and hence $\bar{\alpha}_j$ are some functions of temperature. Thus, if at
some temperature one of the above momentum modes vanishes, we fall at the
point of phase transition. It can be shown that all other modes
$Q^{\,2}(p\,|\,q)$ are definitely positive at all temperatures, there are
no other critical points for physical values of the $\alpha$-parameters.
The possible criticality conditions can be combined into one equation:
$$
\bar{\alpha}_0\,\bar{\alpha}_1\,\bar{\alpha}_2\,\bar{\alpha}_3\;=\;0\,.
\eqno(16)
$$
It can be shown also that if all bond interactions are ferromagnetic
then $\bar{\alpha}_1, \bar{\alpha}_2, \bar{\alpha}_3$ never become zero,
and the critical point can only be associated with $\bar{\alpha}_0=0$, or
$\alpha_0- \alpha_1 -\alpha_2 -\alpha_3 =0$. The criticality conditions
with $\bar{\alpha}_k =0$, $k=1,2,3$, can only be realized when the
antiferromagnetic interactions are involved in the hamiltonian.

If the critical point is associated with $\bar{\alpha}_j=0$, then the
singular part of the free energy (9) near $T_c$ is given as follows [8]:
$$
-\beta f_Q\,|\,_{\rm singular }^{} = \frac{(2\bar{\alpha}_j)^2}
{16\pi\sqrt{(\alpha_0\alpha_1\alpha_2\alpha_3)_c}}
\,\ln\,\frac{\rm const}{(2\bar{\alpha}_j)^2}\;,
\eqno(17)
$$
which implies that near $T_c$ in order of magnitude $-\beta f_Q \sim
\tau^2\,\ln \tau^{\,2}$, where $\tau \sim |T-T_c|$. The specific heat
have thus the log-type singularity, $C \sim |\,\ln\tau\,|$ as $T
\rightarrow T_c$ (Onsager, 1944).

It is seen that the eigenvalues (14) play important role, but it is not so
clear, in fact, how the role of the polynomial $F_{123}$ can be understood
in a less formal way, at the level of the original spin-variable
formulation of the problem, prior to the analytic solution. Even more
amusing grounds to be interesting in this game with the parameters
$\alpha_{j}^{},\, \bar{\alpha}_{j}^{},\, \alpha_{j}^{*}\,$ follow from the
expression for the spontaneous magnetization below $T_c$. The 8-th
power of the spontaneous magnetization $M=<\sigma_{mn}>$ for model (6) is
given by the following very simple expression [8]:
$$
M_{}^{\,8}\,=\,(-1)\,
\frac{\bar{\alpha}_0\bar{\alpha}_1\bar{\alpha}_2\bar{\alpha}_3}
{\alpha_0\alpha_1\alpha_2\alpha_3}\,=\,1\,-\,
\frac{\alpha_{0}^{*}\alpha_{1}^{*}\alpha_{2}^{*}\alpha_{3}^{*}}
{\alpha_0\alpha_1\alpha_2\alpha_3}\,.
\eqno(18)
$$
This expression for $M^{\,8}$ holds true when the right hand side varies
between 0 and 1, and $M^8=0$ otherwise. The well known expressions for the
spontaneous magnetizations of the rectangular, triangular, and hexagonal
lattices follow easily from (18) as particular cases. From (18) we find
$M \sim\tau^{ \frac{1}{8}}$ as $\tau \sim |\,T-T_c\,|\,\to0$, with the
universal value of the critical index $\beta=1/8$ for the magnetization
at the critical isobar. What are the hidden reasons for such simple
expression for $M^{\,8}$, is yet unknown.

{\em 4. Majorana field theory for the 2D Ising models.}

The fermionic integral for $Q$ is a suitable starting point to formulate
the continuum field theory near $T_c$. We write once again the exact
lattice action from (7) as follows:
$$
\ba{llr}
S\,(c)\,= \sum\limits_{mn}\;[\,(\alpha_0-\alpha_1-\alpha_2
-\alpha_3)\,c_{mn}^{}c_{mn}^{*} +\,\alpha_1\,c_{mn}^{}\,(c_{mn}^{*}-
c_{m-1n}^{*})\, + \cr\triex\triex
+\,\alpha_2\,c_{mn}\,(c_{mn}^{*}- c_{mn-1}^{*})
+\,\alpha_3\, c_{mn}\,(c_{mn}^{*}-c_{m-1n-1}^{*})\, +
\cr\triex\triex +\,
\alpha_1\,c_{mn}^{}\,(c_{mn}^{}-c_{m-1n}^{})
+\,\alpha_2\,c_{mn}^{*} (c_{mn}^{*}-c_{mn-1}^{*} )\,]\,,
\cr\ea\eqno(19)
$$
with $\,c_{mn}^{\,2}=c_{mn}^{\,*\;2}=0\,$. Let us define lattice
derivatives $\partial_m\,x_{mn}=x_{mn}-x_{m-1n}\,,\,
\partial_n\,x_{mn}=x_{mn}- x_{mn-1}\,$, with $x_{mn}-x_{m-1n-1}=
\partial_m\,x_{mn} + \partial_n\,x_{mn}-\partial_m \partial_n\,x_{mn}\,$.
Introducing the new notation for the fermionic fields, $c_{mn}\,,\,
c_{mn}^{*} \to \psi_{mn}\,,\, \bar{\psi}_{mn}\,$, from (19) we obtain:
$$
\ba{llr}
S\,(\psi) =\sum\limits_{mn}^{}\,[\;
\underline{m}\,\psi_{mn}\bar{\psi}_{mn} +\,\lambda_1\,\psi_{mn}\,
\partial_m\, \bar{\psi}_{mn}+
\,\lambda_2\,\psi_{mn}\,\partial_n\, \bar{\psi}_{mn}\,-\,
\cr\triex
\,-\,\alpha_3\,\psi_{mn}\,\partial_m\,\partial_n\,\bar{\psi}_{mn} +
\alpha_1\,\psi_{mn}\,\partial_m\,\psi_{mn} +
\alpha_2\,\bar{\psi}_{mn}\, \partial_n\,\bar{\psi}_{mn}\,]\,, \;\;\;
\cr\cr\triex
\lambda_1=\alpha_1+\alpha_3,\;\; \lambda_2=\alpha_2+\alpha_3\,,\;\;\;
\underline{m}=\alpha_0 -\alpha_1 -\alpha_2 -\alpha_3= 2\,\bar{\alpha}_0\,.
\cr\ea\eqno(20)
$$
It is easy to recognize in this still exact lattice action the typical
field-theoretical like structure with mass term and kinetic part.
Remember that we assume the ferromagnetic case with the criticality
condition $\,2\,\bar{ \alpha}_0=0\,$. The parameter $\underline{m}=
\alpha_0- \alpha_1- \alpha_2-\alpha_3 = \,2\,\bar{\alpha}_0\,$ play
the role of mass in the field-theoretical interpretation,
$\,\underline{m} \sim|\,T-T_c\,|\to 0\,$ as $T\to T_c\,$, the ordered
phase corresponds to $\underline{m}<0\,$.

Taking in (20) the formal limit to the continuum euclidean space, with
$(mn)\to (x_1\,|\,x_2)$, $\,\psi_{mn} \to \psi(x)= \psi\,\{\,x_1\,|\,
x_2\,\}$ and $\partial_m \to \partial_1=\partial/\partial x_1$, $\partial_n
\to \partial_2= \partial/\partial x_2\,$, and neglecting the second-order
$\partial_1\partial_2$ kinetic term, we obtain the Majorana like fermionic
action of the correspondent continuum theory:
$$
\ba{llr}
S = \int d^2x \;[\,\underline{m}\,\psi(x)\ovl{\psi}(x) +
\lambda_1\, \psi(x)\,\partial_1\,\ovl{\psi}(x) +
\lambda_2\, \psi(x)\,\partial_2\,\ovl{\psi}(x)\,+
\cr\cr\triex\triex  +\,
\alpha_1\, \psi(x)\,\partial_1\,\psi(x) +
\alpha_2\, \ovl{\psi}(x)\,\partial_2\,\ovl{\psi}(x)\,]\,,
\cr\ea\eqno(21)
$$
where $\lambda_1$, $\lambda_2$, $\underline{m}$ are defined in (20). A
nonstandard feature in this action is the presence of the "diagonal"
kinetic terms like $\psi\partial_1\bar{\psi}$, $\psi\partial_2 \bar{\psi}$.
This terms can be eliminated by a suitable linear transformation of the
fields $\psi,\bar{\psi}$. Omitting the details, after a suitable
transformation of the fields and the $d^{\,2}x$ space in (21), we come to
the euclidean 2D Majorana action in canonical form:
$$
\ba{ccc}
S = \int d^{\,2}x\;[\,i\,\ovl{m}\, \psi_1(x)\psi_2(x) +\,
\psi_1(x)\,\pal_0\,\psi_1(x) + \psi_2(x)\,\ovl{\pal}_0\,\psi_2(x)\,]\,,
\cr\cr
\pal_0= \frac{1}{2}\,(\pal_1-i\pal_2)\,,\;\;\ovl{\pal}_0=
\frac{1}{2}\,(\pal_1+i\pal_2)\,, \;\;
\mbox{2D Majorana}\,,
\ea\eqno(22)
$$
with the rescaled mass:
$$
\ovl{m}= \frac{\alpha_0 -\alpha_1-\alpha_2 -\alpha_3}
{\left(2\,\sqrt{(\alpha_0\alpha_1\alpha_2
\alpha_3)_{c}}\,\right)^{\,1/2}}\;.
\eqno(23)
$$
The 2D Ising model is presented in (22) as a field theory of free massive
two-component Majorana fermions over the euclidean $d^{\,2}x$ space. The
Majorana fields $\,\psi_1\,,\,\psi_2\,$ in this representation are the
linearly transformed fields $\,\psi\,,\,\bar{\psi}\,$ from (21). The axis
of the $d^{\,2}x$ space are also rescaled and rotated as we pass from (21)
to (22). Respectively, the mass is rescaled according (23), here $(\,)_c$
means the criticality condition $(\alpha_0- \alpha_1 -\alpha_2-\alpha_3)_c
=0$. The Majorana field theory for the 2D Ising model, for the simplest
case of the isotropic rectangular lattice, was constructed, by another
method, in refs. [3,4]. The action (22) is a generalization of eq. (129) in
[4], also see eqs. (2.11) in the first reference of [3]. The generalized
action (22) captures the relevant features of the exact lattice 2DIM
theories for the lattices covered by (6), see also (7), in the low-momentum
sector, which is responsible for the large-distance behavior of the
correlation functions and the thermodynamic singularities near $T_c$. In
particular, the result (17) can still be recovered from (22).

{\em 5. The Dirac field interpretation of the 2D Ising models.}

We can pass as well to the Dirac field theory of charged fermions by
doubling the number of fermions in the Majorana representation (22). On
formal level this corresponds to passing from Pfaffian like Gaussian
integral to the determinantal Gaussian integral according the identity
$|\Pfaff\,A\,| ^{\,2}=\det A$, see [9]. To this end, we take two identical
copies $S'$ and $S''$ of (22) and consider the combined action
$S_{\rm\,dirac}=S'+S''$. In this action we introduce the new
fermionic fields $\psi_{1}, \psi_{2}, \psi_{1}^{\,*}, \psi_{2}^{\,*}$ by
substitution:
\begin{small} $$
\ba{llr}
\psi_{1}^{\,'}\,= \frac{1}{\sqrt{2}}(\psi_1+ \psi_{2}^{\,*})\,, \;
i\,\psi_{1}^{\,''}\,= \frac{1}{\sqrt{2}}(\psi_1- \psi_{2}^{\,*})\,, \;
\psi_{2}^{\,'} = \frac{1}{\sqrt{2}}(\psi_2+ \psi_{1}^{\,*})\,, \;
i\,\psi_{2}^{\,''} = \frac{1}{\sqrt{2}}(\psi_2- \psi_{1}^{\,*})\,.
\ea \eqno(24)
$$ \end{small}
In terms of new fields the action $S=S^{'}+S^{''}$ becomes:
$$
\ba{llr}
S_{\,\rm\,dirac}= \int d^{\,2}x\;[\,i\,\overline{m}\;
(\,\psi_1(x)\,\psi_{1}^{*}(x)\,-\psi_2(x)\,\psi_{2}^{*}(x)) +
\triex \cr\cr +\,
\psi_1(x)\,(\pal_1-i\pal_2)\,\psi_{2}^{*}(x) +
\psi_2(x)\,(\pal_1+i\pal_2)\,\psi_{1}^{*}(x)\;]\,,
\ea\eqno(25)
$$
where the rescaled mass $\ovl{m}$ is given in (23).
In matrix notation we find:
$$
\ba{llr}
S= \int d^{\,2}x\;\left\{\,
\left(\ba{c} \psi_{1}^{} \cr \psi_{2}^{} \ea\right)
\left[\, i\,\ovl{m}\left(\ba{lr} 1 & 0 \cr 0 & -1 \ea\right) +
\pal_1\left(\ba{lr} 0&1\cr 1&0\ea\right) +
\pal_2\left(\ba{lr}0&-i \cr i&0\ea\right)
\right]\left(\ba{c} \psi_{1}^{\,*} \cr \psi_{2}^{*\,}
\ea\right)\right\}\,.
\ea\eqno(26)
$$
Introducing the 2D euclidean $\gamma$-matrices $\gamma_1= \sigma_1$,
$\gamma_2= \sigma_2$ and $\gamma_5= \gamma_1\gamma_2= i\sigma_3$, where
$\sigma_1, \sigma_2, \sigma_3$ are standard Pauli matrices, we obtain the
same action in the $\gamma$-matrix interpretation:
$$
\ba{ccc}
S= \int d^{\,2}x\;[\, \Psi^{}\,(x)\,(\,\overline{m}\,
\gamma_5 + \gamma_1\,\pal_1 +
\gamma_2\,\pal_2\,)\,\Psi^{\,\dagger}\,(x)\,]\;\;\;
\mbox{ (euclidean 2D Dirac)}\,, \cr\cr
\gamma_\mu\gamma_\nu + \gamma_\nu\gamma_\mu=2\,\delta_{\mu\nu}\,,\;
\gamma_{1}^{2}=\gamma_{2}^{2}=1\,, \;
\gamma_5=\gamma_1\gamma_2, \;
\gamma_{5}^{2}= -1\,,
\ea\eqno(27)
$$
where $\Psi= (\psi_1\,|\,\psi_2)$, $\Psi^{\,\dagger}= (\psi_{1}^{\,*}\,|\,
\psi_{2}^{\,*})$ are the two-component charged spinors. We find here
the euclidean action of the massive free 2D Dirac theory. The
unconventional $\gamma_5$ matrix in the mass term can be eliminated, if one
wants, by a redefinition of the fields, changing the sign of a one of the
four spinor components.

We can pass as well to the 2D Minkovsky space. The 2D Minkovsky space is
the complex plane. Hence we put $\{\,x_1\,|\,x_2\,\} \rightarrow
\{\,x\,|\,i\,t\}\,,\;\; \Psi\,(x_1\,|\,x_2) \rightarrow
\Psi\,(x\,|\,it)\,$, $\,\pal_1\to \pal_1=\pal/\pal x\,, \;\; \pal_2
\to -\,i\, \pal_0= -i\,\pal/\pal t$. From (25) or (26) we obtain:
$$
\ba{llr}
S_{\,\rm dirac}=
\,i\,\int d^{\,2}x\;\Psi^{}\,(x)\,(\sigma_3)\;[\,\overline{m}\,
+\,\pal_1\,(\sigma_2)\,+\,\pal_0\,(\,i\sigma_1)\,]\,
\Psi^{\,\dagger}\,(x)\,.
\cr\ea\eqno(28)
$$
Introducing the Minkovsky-space $\gamma$-matrices $\gamma^{\,0}=
\sigma_1$, $\gamma^{\,1} =-i\sigma_2$, $\gamma^{\,5}= \gamma^{\,0}
\gamma^{\,1}=\sigma_3$, and passing to new spinors  $\bar{\Psi}=
\Psi\gamma^{5}\,, \; \bar{\Psi}^{\dagger}= \Psi^{\dagger}$, we find:
$$
\ba{llr}
S=\,i\,\int d^{\,2}x\;\bar{\Psi}\,(x)\,[\,\overline{m}\,+
\,i\,\hat{\partial}\,]\;\bar{\Psi}^{\,\dagger}\,(x)\;\;\;
\mbox{(minkovsky 2D Dirac)}\,,
\cr\cr
\hat{\partial}=
\gamma^{\,\mu}\pal_{\,\mu} = \,\gamma^0\,\pal_0 + \gamma^1\,\pal_1\,,\;
\gamma^{\,\mu}\gamma^{\,\nu} + \gamma^{\,\nu}\gamma^{\,\mu}=
2\,g^{\,\mu\nu}\,,\;\, g^{\,\mu\nu} = \mbox{diag}\,(+\,|\,-)\,.
\cr\ea\eqno(29)
$$
We have here the 2D Dirac action in the Minkovsky space interpretation.
This is to be compared with the action given in eq.~(3.17) in [6] for
a particular case of the rectangular lattice.

In conclusion, we have discussed the fermionic structure and related
aspects of the 2D Ising models described by the three-spin polynomial
partition function (6) and the equivalent Gaussian fermionic integral (7).
The symmetries and the criticality conditions are most naturally expressed
in terms of the parameters $\alpha$, $\bar{\alpha}$, $\alpha^{\,*}$ arising
in this spin-polynomial interpretation. Also, the spontaneous magnetization
is expressible very simply in terms of these parameters, see (18), which
property is not yet well understood. We then derived, directly from the
exact lattice action, the continuum-limit Majorana and Dirac field theories
for the models under discussion, including the standard rectangular,
triangular, and hexagonal 2DIM lattices. We find that the lattice
parameters are merely absorbed, in the continuum limit, in the definition
of mass. This confirms the universality ideas. Finally, we argue that even
the standard rectangular lattice can be most easily analyzed, both in the
exact lattice theory and in the field-theoretical approximation, starting
with the spin-polynomial interpretation of the problem.
\end{document}